\documentclass[fleqn,10pt]{wlscirep}
\usepackage[utf8]{inputenc}
\usepackage[T1]{fontenc}
 \usepackage{tcolorbox}

\title{Cyber-Physical Defense in the Quantum Era}

\author[1]{Michel Barbeau*}
\author[2]{Joaquin Garcia-Alfaro*}
\affil[1]{Carleton University, School of Computer Science, Canada.}
\affil[2]{Institut Polytechnique de Paris, T\'el\'ecom SudParis, France.}

\affil[*]{Correspondence to barbeau@scs.carleton.ca and garcia\_a@telecom-sudparis.eu}

\usepackage{soul}
\usepackage{subfigure,wrapfig}
\usepackage{braket}
\usepackage{comment}
\usepackage{glossaries}
\newacronym{api}{API}{Application Programming Interface}
\newacronym{ai}{AI}{Artificial Intelligence}
\newacronym{cpss}{CPSs}{Cyber-Physical Systems}
\newacronym{cps}{CPS}{Cyber-Physical System}
\newacronym{dl}{QD}{Quantum Defense}
\newacronym{dos}{DoS}{Denial of Service}
\newacronym{scada}{SCADA}{Supervisory Control and Data Acquisition}
\newacronym{ict}{ICT}{Information and Communications Technology}
\newacronym{lti}{LTI}{Linear Time Invariant}
\newacronym{lqg}{LQG}{Linear Quadratic Gaussian}
\newacronym{lqr}{LQR}{Linear Quadratic Regulator}
\newacronym{mdp}{MDP}{Markov Decision Process}
\newacronym{cdf}{CDF}{Cumulative Distribution Function}
\newacronym{ml}{ML}{Machine Learning}
\newacronym{mtd}{MTD}{Moving Target Defense}
\newacronym{ncss}{NCSs}{Networked-Control Systems}
\newacronym{ncs}{NCS}{Networked-Control System}
\newacronym{rl}{RL}{Reinforcement Learning}
\newacronym{sdn}{SDN}{Software Defined Networking}
\newacronym{qml}{QML}{Quantum Machine Learning}

\usepackage{siunitx}
\input{Qcircuit}

\begin{abstract}
Networked-Control Systems (NCSs), a type of cyber-physical systems, consist of
tightly integrated computing, communication and control technologies.
While being very flexible environments, they are vulnerable to
computing and networking attacks. Recent NCSs hacking incidents
had major impact. They call for more research on cyber-physical
security. Fears about the use of quantum computing to break current
cryptosystems  make matters worse. While the quantum threat 
motivated the creation of new disciplines to handle the issue, such as
post-quantum cryptography, other fields have overlooked the existence of
quantum-enabled adversaries. This is the case of cyber-physical
defense research, a distinct but complementary discipline to
cyber-physical protection. Cyber-physical defense refers to the
capability to detect and react in response to cyber-physical attacks.
Concretely, it involves the integration of mechanisms to identify
adverse events and prepare response plans, during and after incidents
occur. In this paper, we make the assumption that the eventually available quantum
computer will provide an advantage to adversaries against defenders, unless they also adopt this technology.
We envision the necessity for a paradigm shift, where an increase of
adversarial resources because of quantum supremacy does not translate
into higher likelihood of disruptions. Consistently with
current system design practices in other areas, such as the use
of artificial intelligence for the reinforcement of attack detection
tools, we outline a vision for next generation cyber-physical defense
layers leveraging ideas from quantum computing and machine learning. Through
an example, we show that defenders of NCSs can learn and
improve their strategies to anticipate and recover from attacks.\\

\medskip

\end{abstract}
\begin{document}

\flushbottom
\maketitle
\thispagestyle{empty}

\section{Introduction}
\label{sec:intro}

\gls*{ncss} integrate computation, communications and physical
processes. Their design involves fields such as computer science,
automatic control, networking and distributed systems. Physical
resources are orchestrated building upon concepts and technologies
from these domains. In a \gls*{ncs}, the focus is on remote control,
which means steering at distance a dynamical system according to
requirements. Determined
according to a target behavior, feedback and corrective control actions are transported over a communication
network.

In a \gls*{ncs}, networks and systems, sometimes called the plant, represent
observable and controllable physical resources. The sensors correspond
to observation apparatus. The actuators represent an abstraction of
devices enabling the control of the networked system. 
Using signals produced by the sensors,
the controller generates commands to the actuators. The coupling of
the controller with actuators and sensors happens through a
communications network. In contrast to a classical feedback-control
system, \gls*{ncss} provide remote control.

\gls*{ncss} are flexible, but vulnerable to computer and network
attacks. Adversaries build upon their knowledge about
dynamics, feedback predictability and countermeasures,
to perpetrate attacks with severe implications~\cite{ding2020,ge2020,ding2018}.
When industrial systems and national infrastructures are victimized,
consequences are catastrophic for businesses, governments and society~\cite{Courtney2021}.
A growing number of incidents have been documented.
Representative instances are listed in Box~1.

Attacks can be looked into from several point of views~\cite{teixeira2015}.
We can consider attacks in relation to an adversary knowledge about a
system and its defenses. 
In addition, we can consider attacks with respect to the criticality of disrupted resources.
For example, a \gls*{dos} attack targeting an element that is crucial to
operation~\cite{Zhu2020}.
Besides, we can take into account the ability of an adversary to analyze signals, such as sensor outputs.
This may enable sophisticated attacks impacting
system integrity or availability. Moreover, there are incidents caused
by human adversarial actions.
For instance, they may forge feedback for disruption purposes. 
\gls*{ncss} must be capable of handling security beyond breach. In
other words, they must assume that cyber-physical attacks will happen.
They should be equipped with cyber-physical defense tools. For instance,
response management tools must assure that crucial operational
functionality are be properly accomplished and cannot be
stopped.
For example, the cooling service of a  nuclear plant reactor or
safety control of an autonomous navigation system are crucial 
functionalities. Other less
important functionalities may be temporarily stopped or partially
completed, such as a printing service. It is paramount to assure that
 defensive tools provide appropriate responses, to rapidly
take back control when incidents occur.

That being said,
the quantum paradigm will render obsolete a number of
cyber-physical security technologies. Solutions that are assumed to be
robust today will be deprecated by quantum-enabled adversaries.
For example, adversaries that will capable of brute-forcing and taking advantage of the upcoming quantum computing power.   Disciplines, such as
cryptography, are addressing this issue. Novel post-quantum
cryptosystems are facing the quantum threat. Other fields, however, have
overlooked the eventual existence of quantum-enabled adversaries. Cyber-physical
defense, a discipline complementary to
cryptography, is a proper example. It  uses artificial intelligence mainly to detect anomalies and anticipate
adversaries.
Hence, it enables \gls*{ncss} with capabilities to
detect and react in response to cyber-physical attacks. More concretely, it
involves the integration of machine learning to identify
adverse events and prepare response plans, while and after incidents
occur. An interesting question is the following.
How a defender will face a quantum-enabled adversary? How can a defender
use the quantum advantage to anticipate  response plans? 
How to ensure cyber-physical defense in the quantum era?
In this
paper, we investigate these questions. We develop foundations of a quantum machine learning defense framework. Through an illustrative example, we show
that a defender can leverage quantum machine learning to address the
quantum challenge. We also highlight some recent methodological and
technological progress in the domain and remaining issues.\\

\begin{tcolorbox}[colback=white!5!white,
                  colframe=white!75!black,
                  title=Box 1 - Representative cyber-physical attacks documented in the media.
                 ]

\begin{tabular}{ p{7.95cm} | p{7.95cm}}
\textbf{Espionage and sabotage of critical facilities}, such as US data breach in 2021 due to the \href{https://www.csoonline.com/article/3600893/fireeye-breach-explained-how-worried-should-you-be.html}{SolarWinds attack} or attempts of \href{https://www.reuters.com/article/us-saudi-aramco-security-idUSKBN2002N2}{Saudi Aramco cyber-sabotage} of oil-processing facilities in 2020.
Similar problems are spanning worldwide.
&

\textbf{Adversarial actions in this scenario}, include USB
injection of corrupted software binaries, drive-by-download malware
installation, spear phishing-based design of websites, and traditional
social engineering manipulation of critical infrastructure employees.\\
~~\\
\hline

\medskip

\textbf{Remote control of navigation systems}, including successful
hacking of \href{http://j.mp/2yW61Dr}{autonomous cars} and \href{http://j.mp/2B1RtrR}{avionic systems}. Studies and general
concern started with a malware that infected over sixty thousand
computers of an \href{http://j.mp/2jaM6uM}{Iranian nuclear facility}. &

\textbf{\newline Adversarial actions} include the use of infection vectors (e.g., USB drives), corrupted updates and patches, radio frequency jamming, radio frequency spoofing, and software binary manipulations.\\

~~\\
\hline
\medskip

\textbf{Disruptions of large-scale industries} have been appointed by
the Federal Office for Information Security of Germany as a serious
concern to European factory and industrial markets. Similar threats
affect~\href{http://bit.ly/2LMqR3H}{drones and smart cities}, as well.&

\textbf{\newline Adversarial actions} include the use of GNSS (Global Navigation Satellite Systems) attacks, e.g., jamming of signals, spoofing and hijacking of communications to downgrade communications to insecure modes (e.g., from encrypted to plain-text communications).\\
~~\\
\end{tabular}

\end{tcolorbox}

The remaining sections are organized as follows. 
Section~\ref{sec:related-work} reviews related work. 
Section~\ref{sec:survey} develops our approach, exemplified with a
proof-of-concept.
Section~\ref{sec:discussion} discusses the generalization of the
approach and open problems. Section~\ref{sec:conclusion} concludes the paper.

\section{Related Work}
\label{sec:related-work}

Protection is one of the most important branches of cybersecurity. It mainly relies on the implementation of state-of-the-art cryptographic protocols. They mainly comprise the use of encryption, digital signatures and key agreement. The security of some cryptographic families are based on computational complexity assumptions. For instance, public key cryptography builds upon factorization and discrete logarithm problems. They assume the lack of efficient solutions that break them in polynomial time. However, quantum enabled adversaries can invalidate these assumptions. They put those protocols at risk~\cite{shor:1997,schor:2020}.  At the same time, the availability of quantum computers from research to general purpose applications led to the creation of new cybersecurity disciplines. The most prominent one is Post-Quantum Cryptography (PQC). 
It is a fast growing  research topic aiming to develop new public key cryptosystems resistant to quantum enabled adversaries. 

The core idea of PQC is to design cryptosystems whose security rely on computational problems that cannot be resolved by quantum adversaries in admissible time. Candidate PQC families include  code-based~\cite{mceliece1978public}, hash-based~\cite{merkle1982secrecy}, multivariate~\cite{patarin1996hidden}, lattice-based~\cite{hof98,reg09} and supersingular isogeny-based~\cite{jao2011towards} cryptosystems. Their security is all based on  mathematical problems that are believed to be hard, even with quantum computation and communications resources~\cite{nielsen2002quantum}. Furthermore, PQC has led to new research directions driven by different quantum attacks. For instance, quantum-resistant routing aims at achieving a secure and sustainable quantum-safe Internet~\cite{satoh2020}.

Besides, quantum-enabled adversaries can  disrupt the operation of classical systems. For example, they can jeopardize availability properties by perpetrating brute-force attacks. Solidifying the integrity and security of the quantum Internet is of chief importance. Solutions to these challenges are being developed and published in the quantum security literature using multilevel security stacks.
They involve the combination of quantum and classical security tools~\cite{Iwakoshi2021}. Cybersecurity researchers emphasized
the need for more works on approaches mitigating the impact
of such attacks~\cite{Giraldo2017}. Following their detection, adequate response 
to attacks is a problem that seems to have received little 
attention.
Specially when we are dealing with quantum 
enabled adversaries.  Intrusion detection, leveraging artificial 
intelligence and machine learning, is 
the most representative category of the detection and reaction paradigm. 

\begin{figure}[!b]
\centering
\includegraphics[width=.52\textwidth]{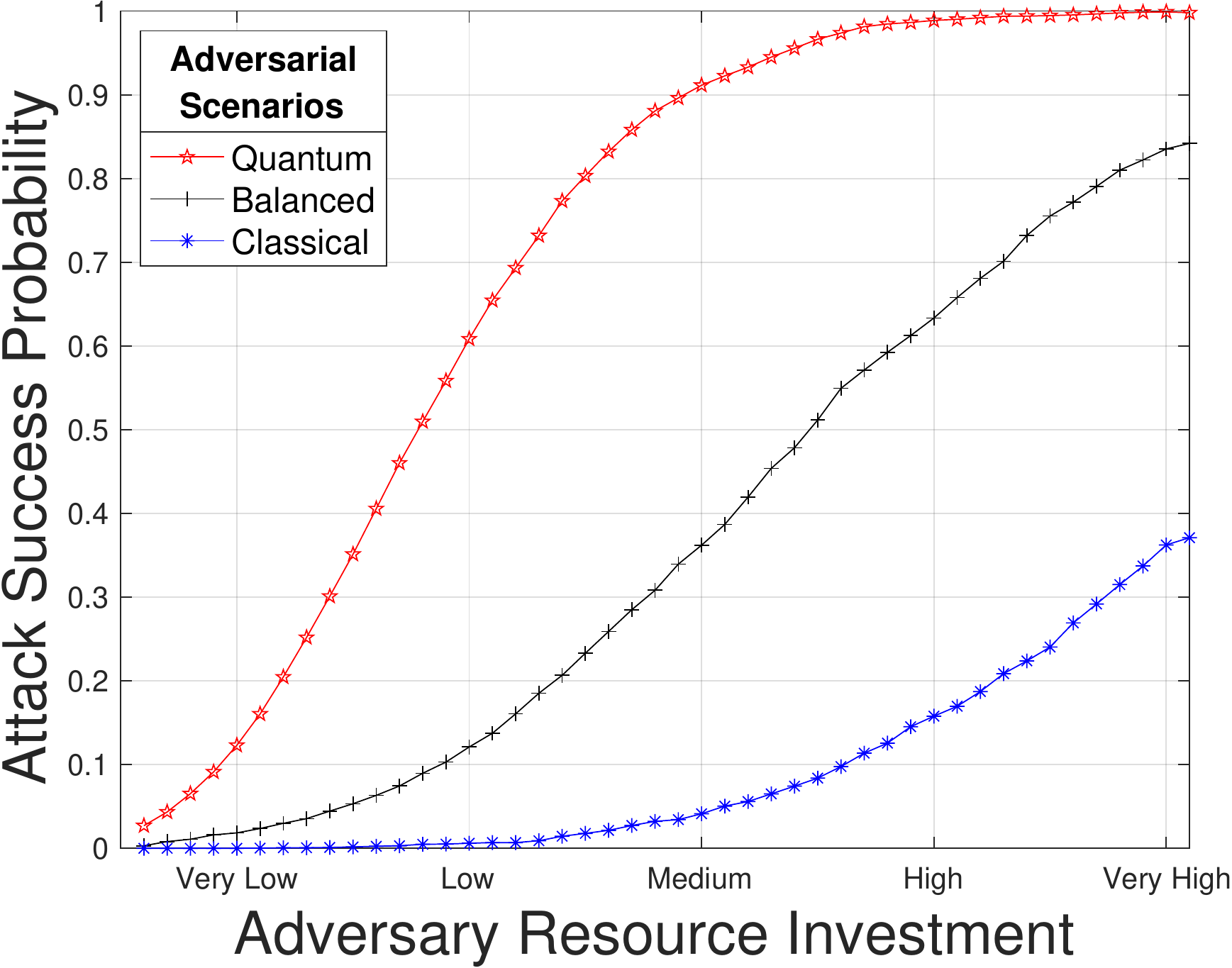}
\caption{Attack success probability vs. adversary investment. We consider three adversarial scenarios. Classical (blue curve), where the resources of the adversary are lower than the resources 
of the defender. Balanced (black), where the resources of the defender are proportional to those of the adversary. Quantum (red), where the resources of the adversary are higher than those of the defender.  Simulation code is available at our companion website, in the Matlab folder~\cite{github}.\label{fig:fig1}}
\end{figure}

The detection and reaction paradigm uses adversarial 
risk analysis methodologies, such as  attack trees~\cite{schneier99} 
and graphs~\cite{lallie2020review}.
Attacks are represented as sequences of incremental steps. The 
last steps of sequences correspond to events detrimental to 
the system. In other words, an attack is considered successful when the
adversary reaches the last step. The cost for the adversary is quantified in terms of resource investment. It is generally assumed that with infinite resources, an adversary reaches an attack probability success of one.
For instance, infinite resources can mean usage of brute-force~\cite{heule2017science}. 
An adversary that increases investment, such as time, computational power or memory, also increases the success probability of reaching the last step of an attack. Simultaneously, this reduces the
likelihood of detection by defenders. Analysis tools may help to explore the relation between
adversary investment and attack success probability~\cite{arnold2014}. Figure~\ref{fig:fig1} schematically depicts the idea.
The horizontal axis represents the cost of the adversary in terms of resource investment. The vertical axis represents the success probability of the attack. We depict three scenarios. 
The blue curve involves a classical adversary with classical resources and a relatively low probability of attack success.
The red curve corresponds to a quantum-enabled adversary, classical defender scenario.
The adversary has the quantum advantage with relatively high probability of attack success.
The black curve represents a balanced situation, where both the adversary and defender have quantum resources.
Every curve models a \gls*{cdf} corresponding the probability of success versus the adversary resource investment. Distribution functions such as
Rayleigh~\cite{rayleigh} and Rician~\cite{gudbjartsson1995rician} are commonly used in the intrusion detection literature for this purpose. Their parameters can be
estimated via empirical penetration testing tools~\cite{arnold2013}.
Without empowering defenders with the same quantum capabilities, an increase of adversarial resources always translate into a higher likelihood of system disruption. In the sequel, we discuss how to equip defenders with quantum resources such that a high attack success probability is not  attainable anymore. 

\section{Cyber-Physical Defense using Quantum Machine Learning}
\label{sec:survey}

\gls*{ml} is about data and, together with clever algorithms, building experience such that next time the system does better.
The relevance of \gls*{ml} to computer security in general has already been
given consideration.
Chio and Freeman~\cite{chio2018machine}
demonstrated general applications of \gls*{ml} to enhance security.
A success story is the use of \gls*{ml} to control spam emails metadata, source reputation, user feedback and pattern recognition are combined to filter out junk emails. Furthermore, there is an evolution capability. The filter gets better with time.
This way of thinking is relevant to \gls*{cps} security because its defense can learn from attacks and make the countermeasures evolve. Focusing on \gls*{cps}-specific threats, as an example pattern recognition can be used to extract from data the characteristics of attacks and to prevent them in the future. Because of its ability to generalize, \gls*{ml} can deal with adversaries hiding by varying the exact form taken by their attacks. Note that perpetrators can adopt as well the \gls*{ml} paradigm to learn defense strategies and evolve attack methods. The full potential of \gls*{ml} for \gls*{cps} security has not been fully explored. The way is open for the application of \gls*{ml} in several scenarios.
Hereafter, we focus on using \gls*{qml} for cyber-physical defense.

\gls*{qml}, i.e., the use of quantum computing for \gls*{ml}~\cite{Biamonte2017}, has potential
because the time complexity of tasks such as classification is independent of the number
of data points. Quantum search techniques are data size independent.
There is also the hope that the quantum computer can learn things that the classical computer is incapable of, due to the fact that the former has properties that the latter does not have, notably entanglement.
At the outset, however, we must admit that a lot remains to be discovered.

\gls*{qml} is mainly building on the traditional quantum circuit model.
Schuld and Killoran investigated the use of kernel
methods~\cite{Schuld2019}, employed for system
identification, for quantum \gls*{ml}.
Encoding of classical
data into a quantum format is involved. A similar approach has been
proposed by Havl{\'\i}{\v c}ek et al.~\cite{havlicek2019}. Schuld and
Petruccione~\cite{schuld2018supervised} discuss in details the application
of quantum \gls*{ml} over classical data generation and quantum
data processing. A translation procedure is required to map the
classical data, i.e., the data points, to quantum data, enabling
quantum data processing, such as quantum classification. However, there
is a cost associated with translating classical data into the quantum
form, which is comparable to the cost of classical \gls*{ml}
classification. This is right now the main barrier. The approach
resulting in real gains is quantum data generation and quantum data
processing, since there is no need to translate from classical to quantum
data. Quantum data generation requires quantum
sensing.

Successful implementation of this approach
will grant a quantum advantage, to the adversary or \gls*{cps} defenders.
There are alternatives to doing \gls*{qml} with traditional quantum circuits.
Use of tensor networks~\cite{montangero2018introduction}, a general graph model, is one of them~\cite{huggins2019towards}. 

Next, we develop an example that illustrates the potential and
current limitations of quantum \gls*{ml}, using variational quantum circuits~\cite{schuld2018supervised,Chen2020,lockwood2020reinforcement},
for solving cyber-physical defense issues.

\subsection{Approach}

\begin{figure}[!b]
\centering
\includegraphics[width=.46\textwidth]{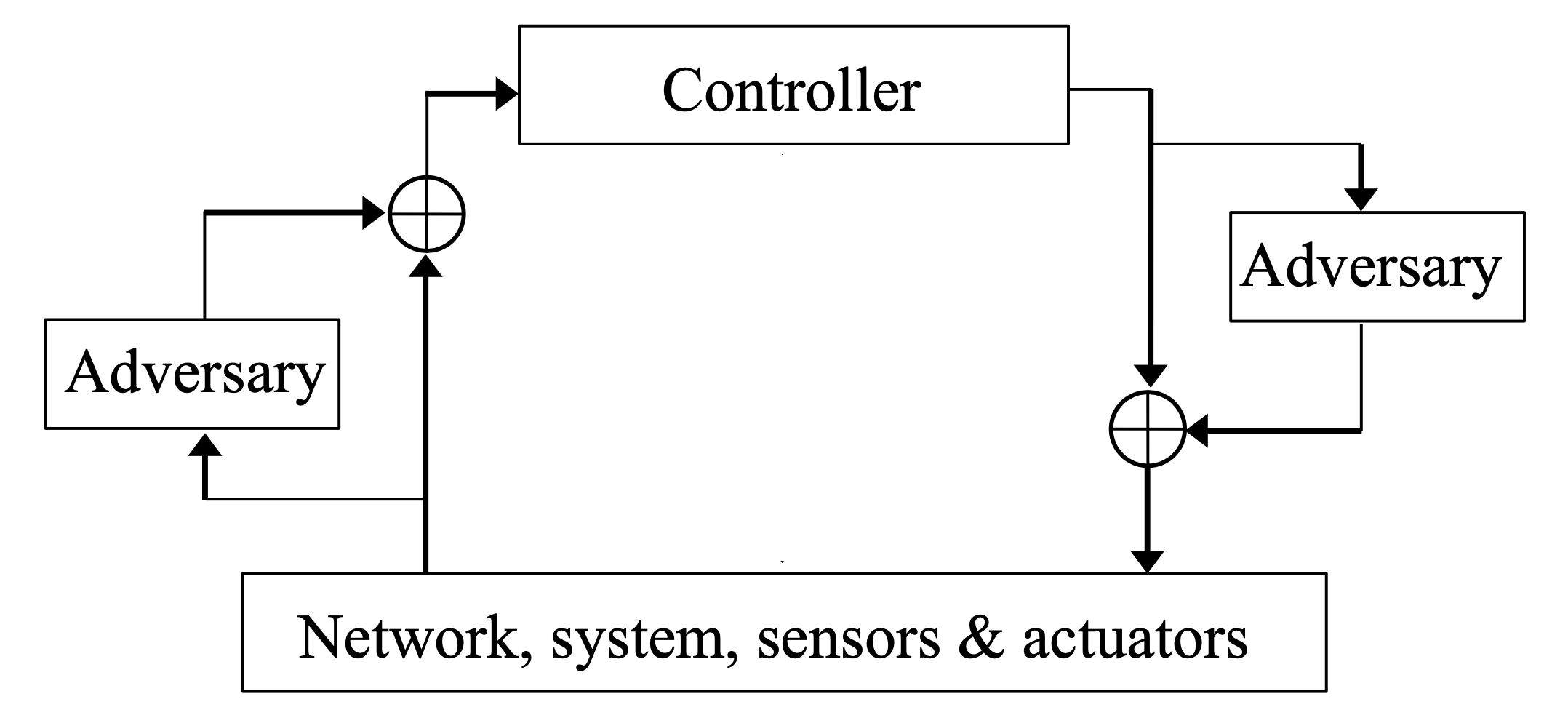}
\caption{Adversarial model.}

\label{fig:fig-advmodel}
\end{figure}

Let us consider the adversarial model represented in
Figure~\ref{fig:fig-advmodel}.
There is a controller getting data and sending control signals through networked sensors and actuators to a system.
An adversary can intercept and tamper signals exchanged between the environment and  controller, in both directions.
Despite the perpetration of attacks, the controller may still have the ability to monitor and steer the system.
This is possible using redundant sensors and actuators attack detection techniques.
This topic has been addressed in related work~\cite{barbeau2021resilience}.
Furthermore, we assume that:
\begin{enumerate}
\item the controller has options and can independently make choices,
\item the adversaries have options and can independently make choices and
\item the consequences of choices made by the controller, in conjunction with those made by adversaries, can be quantified, either by a penalty or a reward.
\end{enumerate}
To capture these three key assumptions, we use the \gls*{mdp} model~\cite{Bellman1957,puterman2014markov}.
The controller is an agent evolving in a world comprising everything else,
including the network, system and adversaries.
At every step of its evolution, the agent makes a choice among a number of available actions, observes the outcome by sensing the state of the world and quantifies the quality of the decision with a numerical score, called reward. Several cyber-physical security and resilience issues lend themselves well to this way of seeing things.

The agent and its world are represented with the \gls*{mdp} model. The quantum learning part builds upon classical \gls*{rl}. The work on \gls*{qml} uses the feature Hilbert spaces of Schuld and Killoran~\cite{Schuld2019}, relying on classical kernel methods.
Classical \gls*{rl},
such as Q-learning~\cite{Watkins1989,Watkins1992},
assumes that the agent, i.e., the learner entity, evolves in a deterministic world. The evolution of the agent and its world is also formally modeled by the \gls*{mdp}.
A \gls*{rl} algorithm trains the agent to make decisions such that a maximum reward is obtained.
\gls*{rl} aims at optimizing the expected return of a \gls*{mdp}.
The objectives are the same with \gls*{qml}.
We explain \gls*{mdp} modeling and the quantum \gls*{rl} part in the sequel.

\subsubsection{MDP Model}

A \gls*{mdp} is a discrete time finite state-transition model that captures random state changes, action-triggered transitions and states-dependent rewards.
A \gls*{mdp} is a four tuple $(S,A,P_a,R_a)$ comprising a set of $n$ states $S=\{ 0, 1,\ldots,n-1 \}$, a set of $m$ actions $A=\{ 0,1,\ldots,m-1 \}$,  a transition probability function $P_a$ and a reward function $R_a$.
The evolution of a \gls*{mdp} is paced by a discrete clock.
At time $t$, the \gls*{mdp} is in state $s_t$, such that $t=0,1,2,\ldots$.
The \gls*{mdp} model starts in an initial state $s_0=0$.
The transition probability function, denoted as
\begin{equation}
P_a(s,s')
=
Pr[ s_{t+1} = s' \vert s_t = s, a_t = a ]
\end{equation}
defines the probability of making a transition to state $ s_{t+1}$ equal to $s'$ at time $t+1$, when at time $t$  the state is $s_t = s$ and action $a$ is performed.
The reward function $R_a(s,s')$ defines the immediate reward associated with the transition from state $s$ to $s'$ and action $a$.
It has domain $S \times S\times A$ and co-domain $\mathbb{R}$.

\begin{tcolorbox}[colback=white!5!white,
                  colframe=white!75!black,
                  title=Box 2 - Quantum computing basics.
                 ]
With quantum computing, the basic unit of information is the quantum bit or qubit.
A qubit is a binary unit of data that is simultaneously a zero and a one until the end of its life when the qubit is measured, which ends either in state zero or one. There is a probability associated with each of these two outcomes. In the ket notation, a qubit is represented by the pair
$$
q=a\ket{0}+b\ket{1}
$$
The symbols $\ket{0}$ and $\ket{1}$, pronounced ket zero and ket one, denote the quantum states zero and one. The parameters $a$ and $b$ are called probability amplitudes. Raised to the power of two, i.e., $a^2$ and $b^2$, they respectively correspond to the probability of measuring the qubit in state zero or state one. The plus sign does not represent arithmetic addition. Rather, an expression with the plus sign should be interpreted as a superposition of its operands, in this case the quantum states $\ket{0}$ and $\ket{1}$. Superposition means that a qubit is both a zero and a one at the same time.
Quantum computations are done by gates.
For instance the Hadamard find many applications.
It can calculate the arithmetic sum of the probability amplitudes $a+b$ and their difference $a-b$.

\tcblower

Several qubits can be grouped together to represent a complex problem. For instance, a two-qubit quantum state $q_1 q_0$, where $q_1$ is equal to $a_1\ket{0}+b_1\ket{1}$ and $q_0$ is equal to $a_0\ket{0}+b_0\ket{1}$, corresponds to the superposition:
$$
a_1a_0\ket{00}+a_1b_0\ket{01}+b_1a_0\ket{10}+b_1b_0\ket{11}
$$
Interestingly, information can be contained in binary combinations but also in probability amplitudes of ket terms. Quantum machine learning leverages both forms of information representation.

\medskip 

Qubits may be entangled, that is, related together such that they read in a coherent way. This means that some of the reading outcomes are made more probable than others. Besides, some reading outcomes can be made entirely non-probable.
\end{tcolorbox}

\begin{tcolorbox}[colback=white!5!white,
                  colframe=white!75!black,
                  title=Box 3 - Bloch sphere representation of a qubit.
                 ]
\begin{wrapfigure}{r}{5cm}
\includegraphics[width=5cm]{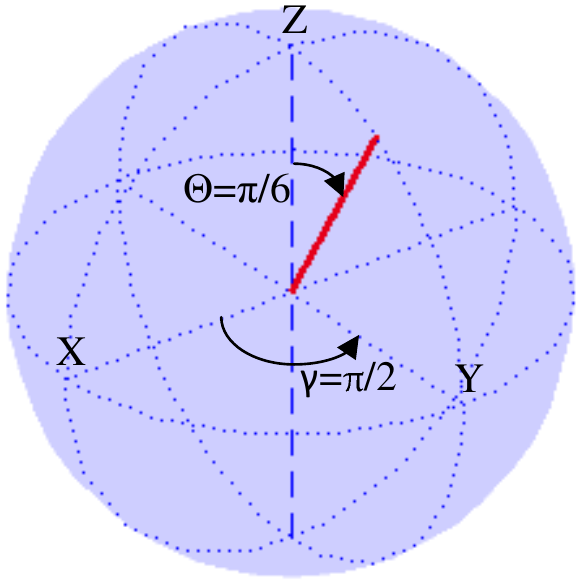}
\end{wrapfigure}
A qubit can be graphically represented as a point on the surface of a unit radius Bloch sphere\footnote{Produced using the Hydrogenic Wavefunction Visulization Tool.}.
In this example, the complementary angle of the latitude, i.e., the colatitude $\theta$ is equal to $\pi/6$. The longitude $\gamma$ is equal to $\pi/2$. A radial line is drawn from the origin to the point on the sphere surface representing the qubit.
The point is defined by spherical coordinates.
The two probability amplitudes are defined by two angles, namely, $\theta$ and $\gamma$.
Angle $\theta$ is the colatitude.
Angle $\gamma$ is the longitude.
In the Bloch sphere framework, a qubit is defined by the superposition
$$
\Ket{\psi}
=
\cos \frac{\theta}{2}
\Ket{0}
+
e^{j\gamma}
\sin \frac{\theta}{2}
\Ket{1}.
$$
With $\theta$ equal to zero, the qubit is $\Ket{0}$.
With $\theta$ equal to $\pi$, the qubit is $\Ket{1}$.
The Bloch sphere representation highlights the difference between a classical bit and a qubit.
A classical bit can only be one of two points on the sphere, the north pole (0) or the south pole (1).
A qubit can be any point on the sphere.
\end{tcolorbox}

\begin{tcolorbox}[colback=white!5!white,
                  colframe=white!75!black,
                  title=Box 4 - Q-learning \gls*{rl}.
                 ]

Q-learning~\cite{Watkins1989} \gls*{rl}
builds upon the idea of associating rewards to actions and states.
A policy, a function $\pi$ with domain state set $S$ and co-domain action set $A$,
associates a recommended action to every state.
Assuming an agent is following a policy $\pi$, every single state $s\in S$  has value
$V_{\pi}(s)$ recursively defined as:
$$
V_{\pi}(s)
=
\sum_{s' \in S} P_{\pi(s)}(s,s') \cdot \left[  R_{\pi(s)}(s,s') + \gamma  V_{\pi}(s')  \right]
$$
After the execution of the policy determined action $\pi(s)$, $s'$ denotes the successor state of $s$.
$P_{\pi(s)}(s,s')$ represents the probability of $s'$ executing action $\pi(s)$.
Under policy $\pi$, the evaluation of $V_{\pi}(s)$ denotes the value of state $s$.
The reward obtained executing action $a$ equal to $\pi(s)$ in state $s$ is $R_a(s,s')$, or $R_{\pi(s)}(s,s')$.
Constant $\gamma$ in $[0,1]$ is a discounting factor, weighting the long-term
reward less than the short term one.
The goal of \gls*{rl} is to find a  policy that makes the agent obtain the best possible reward.
Best possible reward is achieved when the world goes through the most valued states.
The optimal policy is such that for every state $s$
$$
V_{\pi}(s)
=
\max_a \left(
\sum_{s' \in S} P_{a}(s,s') \cdot \left[  R_{a}(s,s') + \gamma  V_{\pi}(s')  \right]
\right).
$$
For obtaining the most rewarding policy,
Q-learning uses the concept of Q-value.
In reference to a policy $\pi$,
it is a function $Q$ with domain $S \cdot A$ and co-domain $\mathbb{R}$, defined as
$$
Q_{\pi}(s,a)
=
\sum_{s' \in S} P_{a}(s,s') \cdot \left[  R_{a}(s,s') + \gamma  V_{\pi}(s')  \right].
$$
The optimization is accomplished through a sequence of epochs $t=0,1,\ldots,n$.
The Q-learning algorithm is, at epoch $t$
for the pair $(s,a)$, where $s\in S$ is the current state and $a\in A$ is the executed action,
$$
\label{eq:bellman}
Q_{t}(s,a)
=
(1 - \alpha) Q_{t-1}(s,a)
+
\alpha \left[ R_a(s,s') + \gamma V_{t-1}(s') \right]
$$
with learning factor $\alpha$ in $[0,1]$.
For every other action pair $(s,a)$,
where $s\in S$ is not the current state or $a\in A$ is not the executed action,
$Q_{t}(s,a)$ is equal to $Q_{t-1}(s,a)$.
At epoch $t-1$, the value of state $s$ is
$$
V_{t-1}(s)
=
\max_a Q_{t-1}(s,a)
$$
For all pairs $(s,a)$, $Q_{0}(s,a)$ is set to null.
It has been established~\cite{Watkins1992} that
$Q_{t}$ converges to the optimal policy when $n$ approaches infinity.
That is, when $t$ tends to infinity we have that $Q_{t}$ tends to $Q_{\pi}$ with the optimal policy $\pi$.
\end{tcolorbox}

\subsubsection{Quantum Reinforcement Learning}\label{sec:qrl}

In this section we present our cyber-physical defense approach.
A reader unfamiliar with quantum computing may first read Boxes~2 and~3, for a short introduction to the topic.
At the heart of the approach is the concept of variational circuit.
Bergholm et al.~\cite{bergholm2020pennylane} interpret such a circuit as the quantum implementation of a function
$f(\psi,\Theta): {\Bbb R}^m\rightarrow {\Bbb R}^n$.
That is, a two argument function from a dimension $m$ real vector space to a dimension $n$ real vector space, where $m$ and $n$ are two positive integers.
The first argument $\psi$ denotes an input quantum state to the variational circuit.
The second argument $\Theta$ is the variable parameter of the variational circuit.
Typically, it is a matrix of real numbers.
During the training,
the numbers in the matrix are progressively tuned, via optimization, such that the
behavior of the variational circuit eventually approaches a target function.
In our cases, this function is the optimal policy $\pi$, in the terminology of Q-learning (see Box 4).

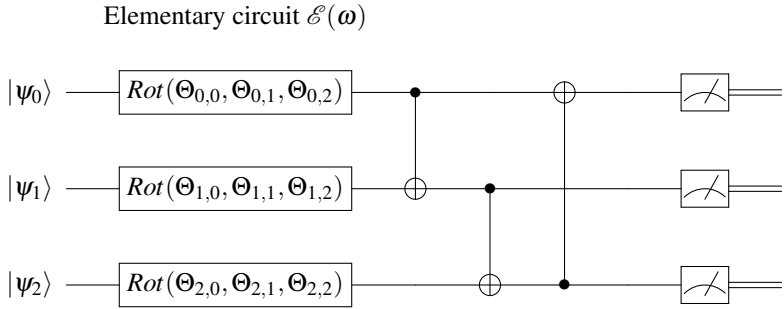
\begin{figure}[htb]
	\begin{center}
		\[
		\Qcircuit @C=2em @R=2em {
			& & \mbox{Elementary circuit $\mathcal{E(\omega)}$} \\
			&\lstick{\ket{\psi_0}} & \gate{Rot(\Theta_{0,0},\Theta_{0,1},\Theta_{0,2})} & \ctrl{1} & \qw      & \targ     & \qw & \meter &  \cw \\
			&\lstick{\ket{\psi_1}} & \gate{Rot(\Theta_{1,0},\Theta_{1,1},\Theta_{1,2})} & \targ    & \ctrl{1} & \qw       & \qw & \meter &  \cw \\
			&\lstick{\ket{\psi_2}} & \gate{Rot(\Theta_{2,0},\Theta_{2,1},\Theta_{2,2})} & \qw      & \targ    & \ctrl{-2} & \qw & \meter &  \cw
		}
		\]
		\caption{Three-qubit variational circuit layer $W(\Theta)$, where $\Theta$ is a three by three matrix of rotation angles.}
		\label{fig:elementarycircuit}
	\end{center}
\end{figure}

As an example, an instance of the variational circuit design of Farhi and
Neven~\cite{Farhi2018} is pictured in Figure~\ref{fig:elementarycircuit}.
In this example, both $m$ and $n$ are three.
It is a $m$-qubit circuit.
A typical variational circuit line comprises three stages: an initial state, a sequence of gates and a measurement device.
In this case, for $i=0,1,2$, the initial state is $\ket{\psi_i}$. The gates are a parameterized rotation and a CNOT.
The measurement device is represented on the extreme right box, with a symbolic measuring dial.
The circuit variable parameter $\Theta$ is a three by three matrix of rotation angles.
For $i=0,1,\ldots,m-1$, the gate
$Rot(\Theta_{i,0},\Theta_{i,1},\Theta_{i,2})$ applies the $x$, $y$ and
$z$-axis rotations  $\Theta_{i,0}$, $\Theta_{i,1}$ and $\Theta_{i,2}$ to qubit $\ket{\psi_i}$ on the Bloch sphere (see Box~3 for an introduction to the Bloch sphere concept).
The three rotations can take qubit $\ket{\psi_i}$ from any state to any state.
To create entanglement between qubits,
qubit with index $i$ is connected to qubit with index $i + 1$, modulo $m$, using a CNOT gate.
A CNOT gate can be interpreted as a controlled XOR operation.
The qubit connected to the solid dot end, of the vertical line, controls the qubit connected to the circle embedding a plus sign.
When the control qubit is one, the controlled qubit is XORed.

In our approach, quantum \gls*{rl} uses and train a variational circuit.
The variational circuit maps quantum states to quantum actions,
or action superpositions.
The output of the variational circuit is a superposition of actions.
During learning, the parameter $\Theta$ of the variational circuit is tuned such that
the output of that variational collapses to actions that are proportional to their goodness, that is, the rewards they provide to the agent.

The training process can be explained in reference to Q-learning.
For a brief introduction to Q-learning, see Box~4.
The variational circuit is a representation of the policy $\pi$.
Let $W(\Theta)$ be the variational circuit.
$W$ is called a variational circuit because it is parameterized with the matrix of rotation angles $\Theta$.
The \gls*{rl} process tunes the rotation angles in $\Theta$.
Given a state $s \in S$, an action $a \in A$ and epoch $t$, the probability of measuring value $a$ in the quantum state $\mathrm{A}$ that is the output of the system
$$
    \mathrm{A} = W(\Theta) \ket{s}
$$
is proportional to the ratio
\begin{equation}\label{eq:propration}
p_{t,s,a}
=
\frac{
  Q_{t}(s,a)
  }{
  \sum_{i \in A} Q_{t}(s,i)}.
\end{equation}
The matrix $\Theta$ is initialized with arbitrary rotations $\Theta_{0}$.
Starting from the initial state $s_0$,
the following procedure is repeatedly executed.
At the $t$\textsuperscript{th} epoch,
random action $a$ is chosen from set $A$.
In current state $s$, the agent applies action $a$ causing the world to make a transition.
World state $s'$ is observed.
Using $a$, $s$ and~$s'$,
the Q-values ($Q_{t}$) are updated.
For every other action pair $(s,a)$,
where $s\in S$ is not the current state or $a\in A$ is not the executed action,
probability $p_{t,s,a}$
is also updated according to Equation~\eqref{eq:propration}.
Using $\Theta_{t-1}$ and the~$p_{t,s,a}$ probabilities,
the variational circuit parameter is updated and yields $\Theta_{t}$.

The variational circuit is trained such that under input state $\ket{s}$,
the measured output in the system $\mathrm{A} = W(\Theta) \ket{s} $ is $a$ with
probability $p_{t,s,a}$.
Training of the circuit can be done with a gradient descent optimizer~\cite{bergholm2020pennylane}.
Step-by-step,
the optimizer minimizes the distance between the probability of measuring $\ket{a}$
and the ratio $p_{t,s,a}$, for $a$ in $A$.

The variational circuit $W(\Theta)$ is trained on the probabilities of the computational basis members of $A$, in a state $s$.
Quantum \gls*{rl} repeatedly updates $\Theta$ such that the
evaluation of $\mathrm{A} =  W(\Theta)\ket{s}$ yields actions with probabilities  proportional to the rewards.
That is, the action action recommended by the policy is $\arg\max_{a\in A} \mathrm{A}$, i.e., the row-index of the element with highest probability amplitude.

Since $W(\Theta)$ is a circuit, once trained it can be used multiple times.
Furthermore, with this scheme the learned knowledge $\Theta$, which are rotations,
can be easily stored or shared with other parties.
This \gls*{rl} scheme can be implemented using the resources of the PennyLane software~\cite{bergholm2020pennylane}.
An illustrative example is discussed in the next subsection.

\subsection{Illustrative Example}\label{sec:example}

In this section, we illustrate our approach with an example.
We model the agent and its world with the \gls*{mdp} model.
We define the attack model.
We explain the quantum representation of the problem.
We demonstrate enhancement of resilience leveraging quantum \gls*{rl}.

\subsubsection{Agent and Its World}

\begin{figure}[!bht]
	\centering
	\subfigure[Configuration \label{fig:two_train}]{
	\includegraphics[width=.45\columnwidth]{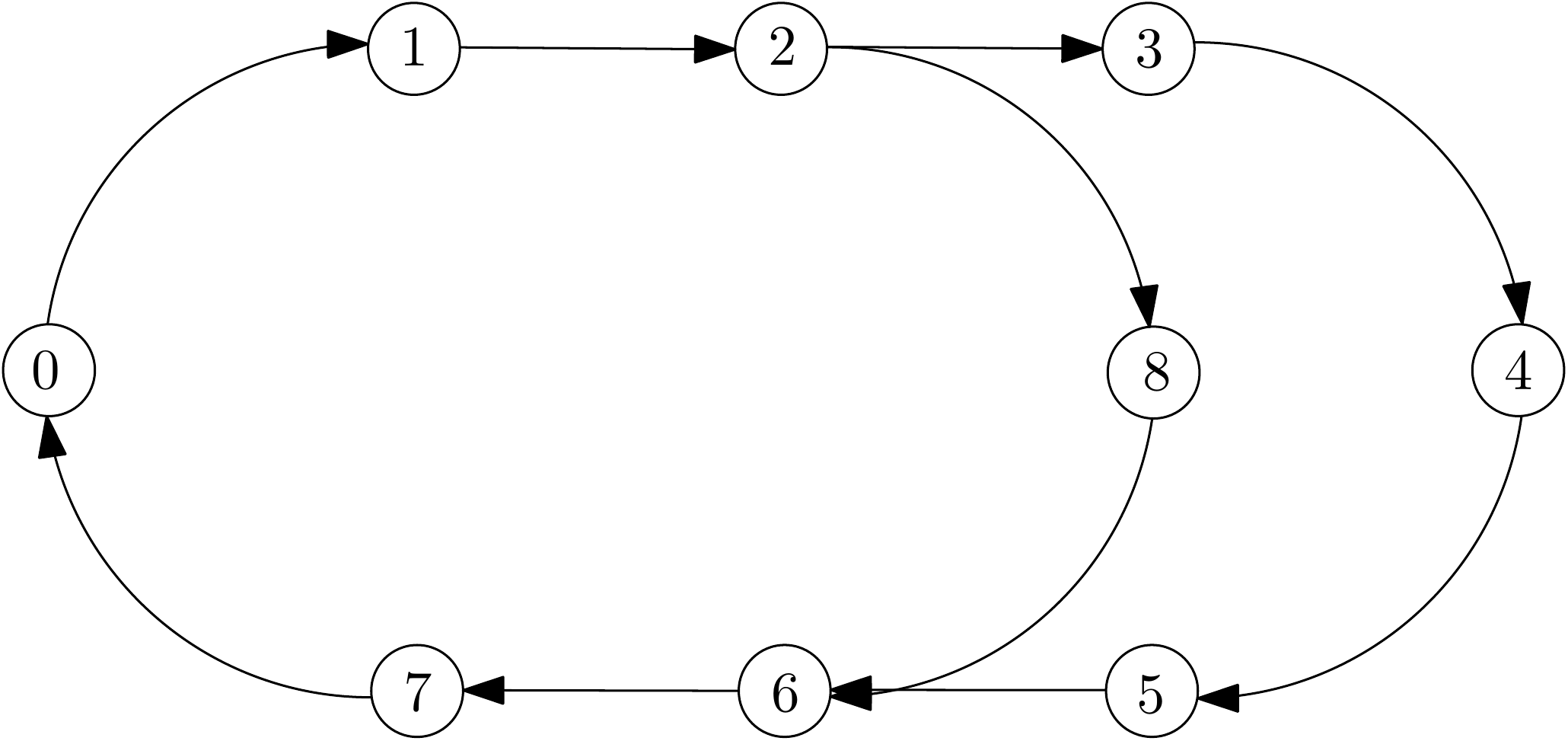}
	}
	\subfigure[MDP \label{fig:mdp}]{
	\includegraphics[width=.35\columnwidth]{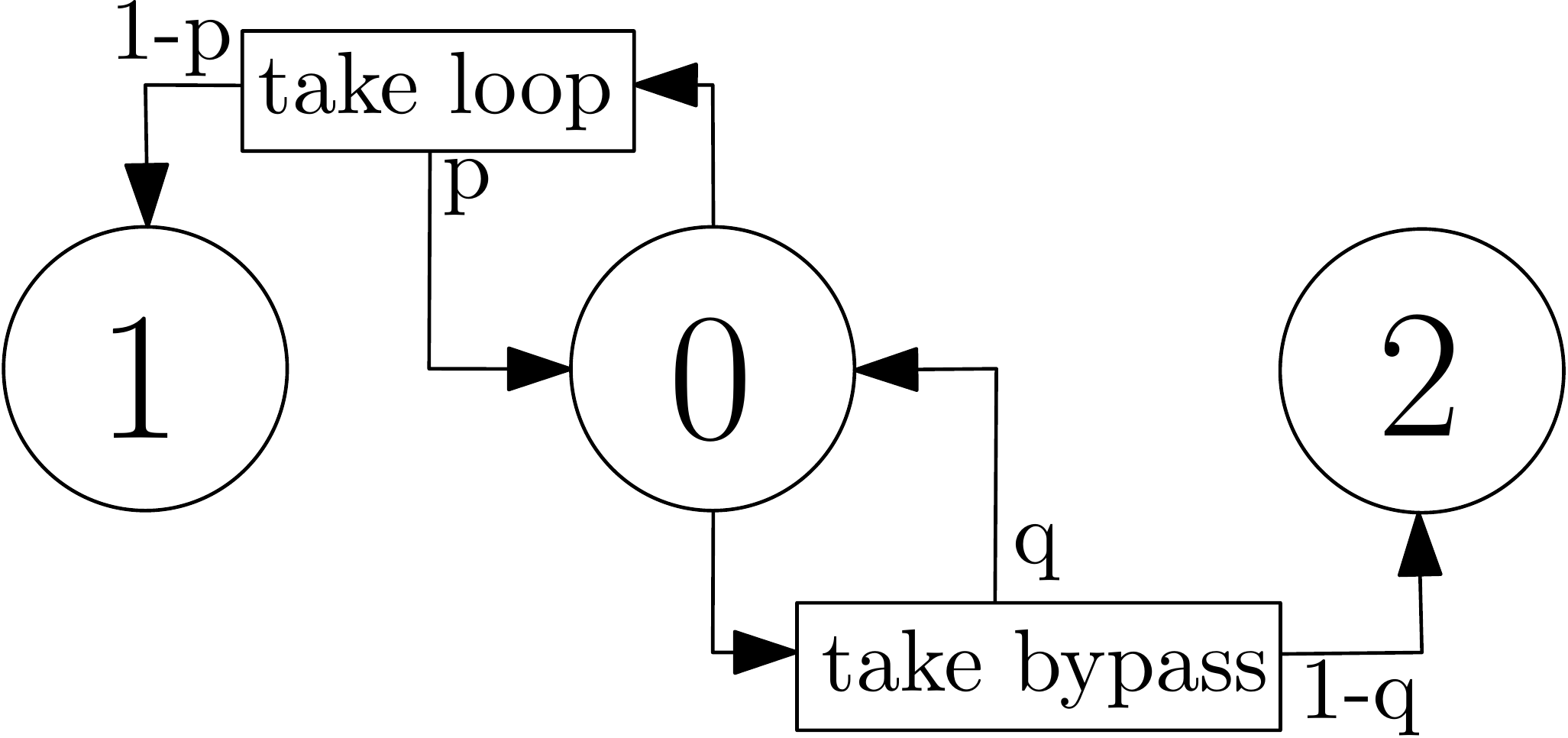}
	}
	\caption{(a) Discrete two-train configuration. (b) MDP representation of the agent in its world. Circles represent states. Arrows represent action-triggered transitions.}\label{fig:ML-trains}
\end{figure}
Let us consider the discrete two-train configuration of Figure~\ref{fig:two_train}.
Tracks are broken into sections.
We assume a scenario where Train~1 is the agent and Train~2 is part of its world.
There is an outer loop visiting points~3, 4 and~5,
together with a bypass from point~2, visiting point~8 to point~6.
Traversal time is uniform across sections.
The normal trajectory of Train~1 is the outer loop,
while maintaining a Train~2-Train~1 distance greater than one empty section.
For example, if Train~1 is at point~0 while
Train~2 is at point~7, then
the separation distance constraint is violated.
The goal of the adversary is to steer the system in a state where the separation distance constraint is violated.
When a train crosses point~0,
it has to make a choice: either traverse the outer loop or take the bypass.
Both trains can follow any path and make independent choices,
when they are at point 0.

In the terms of \gls*{rl},
Train~1 has two actions available: take loop and take bypass.
The agent gets $k$ reward points for a relative Train~2-Train~1 distance increase of $k$ sections with Train~2.
It gets $-k$ reward points, i.e., a penalty,
for a relative Train~2-Train~1 distance decrease of $k$ sections with Train~2.
For example, let us assume that Train~1 is at point~0 and that Train~2 is at point~7.
If both trains, progressing a the same speed, take the loop or both  decide to take the bypass,
then there is no relative distance change.
The agent gets no reward.
When Train~1 decides to take the bypass and Train~2 decides to take the loop,
the agent gets two reward points,
at return to point zero (Train~2 is at point five).
When Train~1 decides to take the loop and Train~2 decides to take the bypass,
the agent gets four reward points,
at return to point zero (Train~2 is at point one, Train~2-Train~1 distance is five sections).

The corresponding \gls*{mdp} model is shown in Figure~\ref{fig:mdp}.
The state set is $S=\{ 0, 1,2 \}$.
The action set is $A=\{ a_0=\mbox{take loop},  a_1=\mbox{take bypass} \}$.
The transition probability function is defined as $P_{a_0}(0,0)=p$,   $P_{a_0}(0,1)=1-p$, $P_{a_1}(0,0)=q$ and $P_{a_1}(0,2)=1-q$.
The reward functions is defined as $R_{a_0}(0,0)=0$, $R_{a_0}(0,1)=4$, $R_{a_1}(0,0)=0$ and $R_{a_1}(0,2)=2$.
This is interpreted as follows.
In the initial state $0$ with a one-section separation distance,
the agent selects an action to perform: take loop or take bypass.
Train~1 performs the selected action.
When selecting take loop,
with probability $p$ the environment goes back to state $0$ (no reward) or with probability $1-p$ it moves to state $1$,
with a five-section separation distance (reward is four).
When selecting take bypass,
with probability $q$ the environment goes back to state $0$ (no reward) or with probability $1-q$ it moves state $2$,
with a three-section separation distance (reward is two).
The agent memorizes how good it has been to perform a selected action.

As shown in this example,
multiple choices might be available in a given state.
A \gls*{mdp} is augmented with a policy.
At any given time, the policy tells the agent which action to pick such that the expected return is maximized.
The objective of \gls*{rl} is finding a policy maximizing the return.
Q-learning captures the optimal policy into a state-action value function $Q(s,a)$,
i.e., an estimate of the expected discounted reward for executing action $a$ in state $s$~\cite{Watkins1989,Watkins1992}.
Q-learning is an iterative process.
$Q_t(s,a)$ is the state-action at the $t$\textsuperscript{th} episode of learning.

\begin{figure}[!htb]
\centering
   \centering
    \subfigure[]{
       \includegraphics[width=0.48\columnwidth]{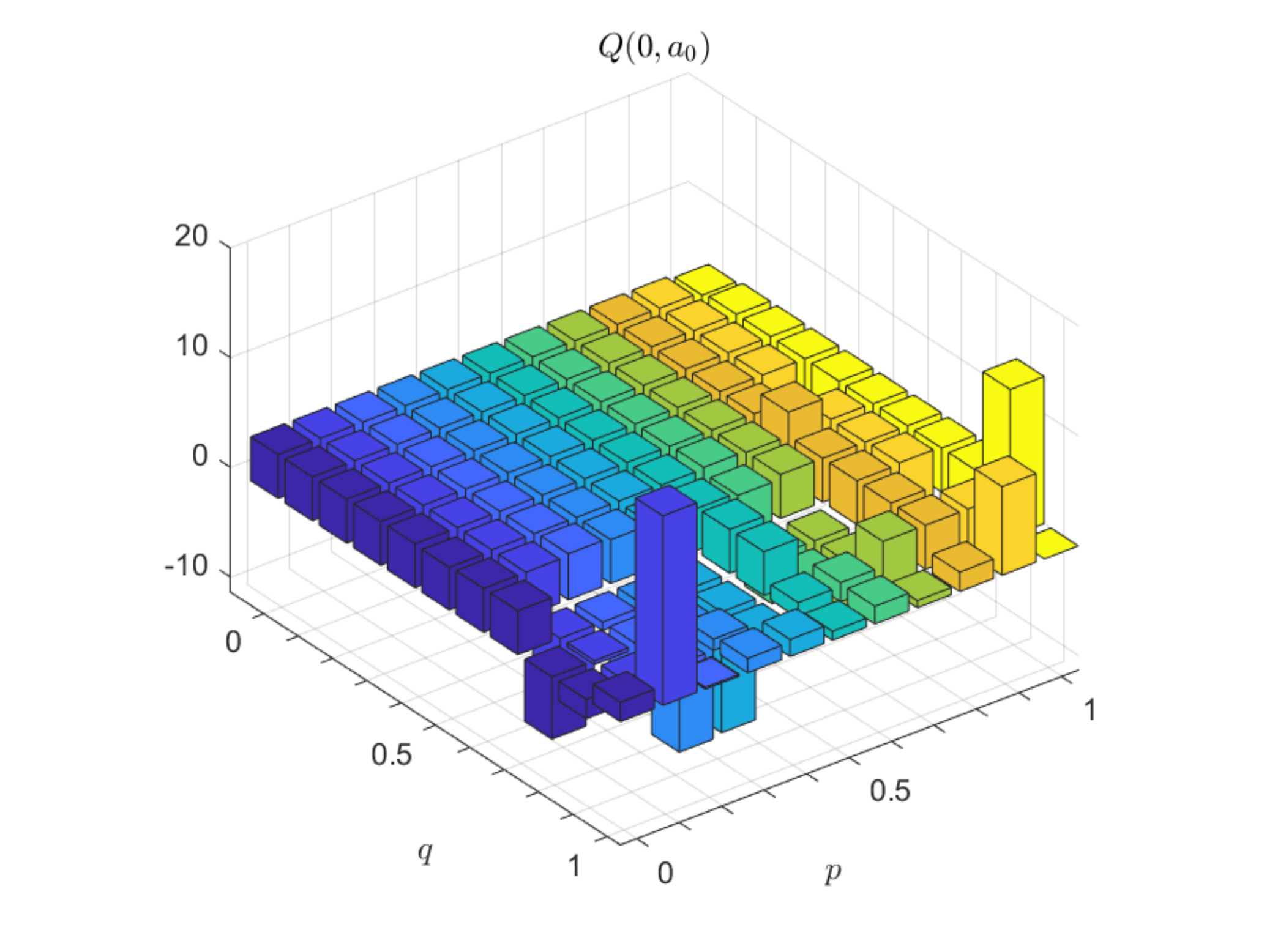}
    }
    \subfigure[]{
       \includegraphics[width=0.48\columnwidth]{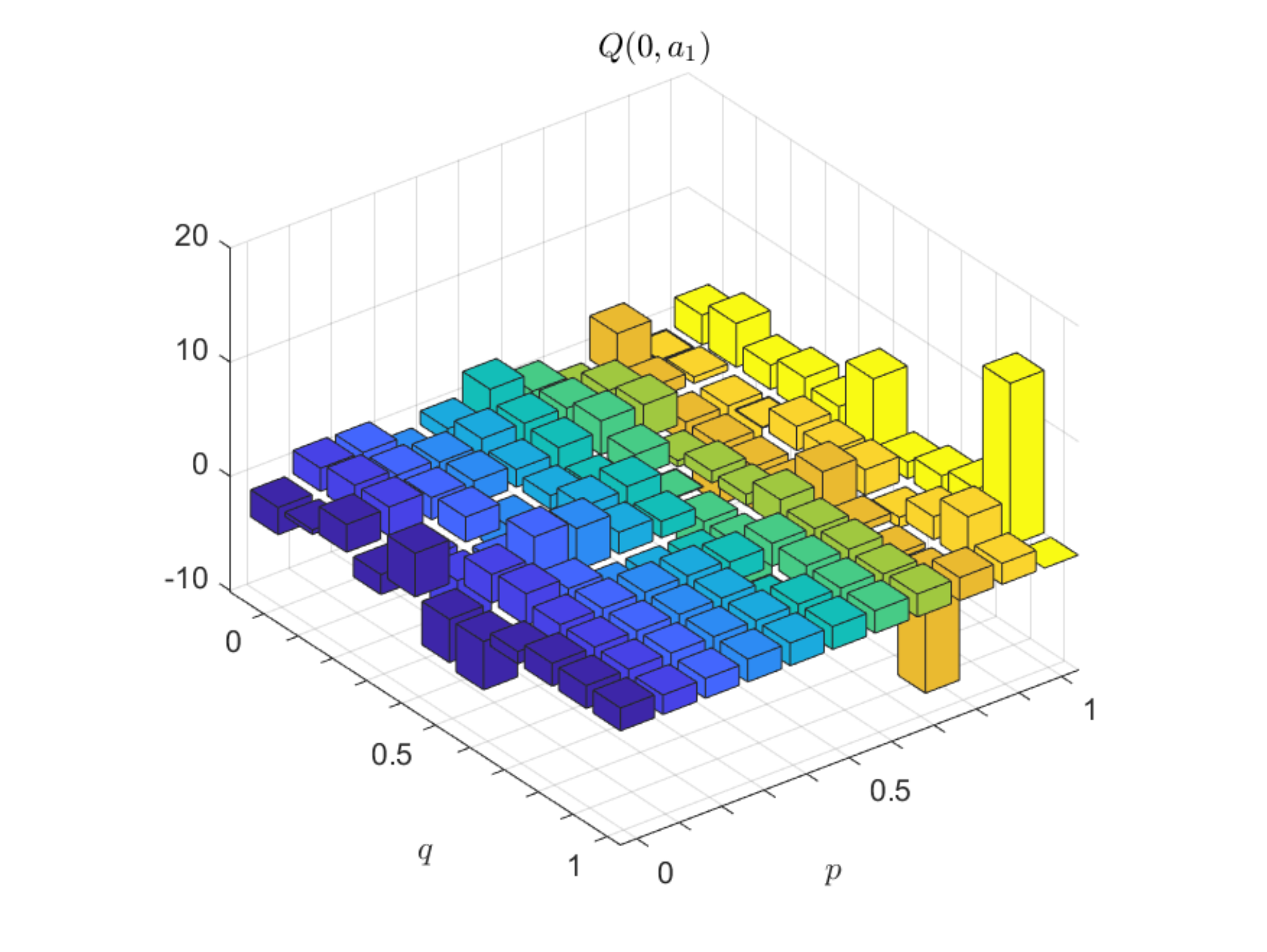}
    }
\caption{Q-values for actions $a_0$ and $a_1$, for values of probabilities $p$ and $q$ ranging from zero to one, in steps of $0.1$.}
\label{fig:q-values}
\end{figure}

Figure~\ref{fig:q-values} plots side by side the Q-values for actions $a_0$ and $a_1$,
for values of probabilities $p$ and $q$ ranging from zero to one, in steps of $0.1$.
As a function of $p$ and $q$, on which the agent has no control, the learned policy is that in state zero should pick the action among $a_0$ and $a_1$ that fields the maximum Q-value, which can be determined from Figure~\ref{fig:q-values}.
This figure highlights the usefulness of \gls*{rl}, even for such a simple example the exact action choice is by far not always obvious.
However, \gls*{rl} tells what this choice should be.

The example is simple enough so that a certain number of cases can be highlighted.
When probabilities $p$ and $q$ tend to one, it means that the adversary is more likely to behave as the agent.
Inversely, when $p$ and $q$ tend to null, the adversary is likely to
make a different choice from that of the agent.
Such a bias can be explained by the existence of an insider that leaks information to the adversary when the agent makes its choice at point~$0$.
In the former case, the agent is trapped in a risky condition.
In the latter case, the adversary is applying its worst possible strategy.
When $p$ and $q$ are both close to 50\%, the adversary is behaving arbitrarily.
On the long term, the most rewarding action for the agent is to take the loop.
It is of course possible to update the policy according to a varying adversarial behavior, i.e., changing values for $p$ and $q$.
In following, we address this \gls*{rl} problem with a quantum approach.\\

\subsubsection{Quantum Representation}
The problem in the illustrative example of Figure~\ref{fig:ML-trains}
comprises only one state ($0$) where choices are available.
A binary decision is taken in that state.
The problem can be solved by a single qubit variational quantum circuit.
The output of the circuit is a single qubit with the simple following interpretation.
$\vert 0 \rangle$ is action take~loop,
while $\vert 1 \rangle$ is action take~bypass.

\begin{figure}[!h]
	\centering
	\includegraphics[width=.49\columnwidth]{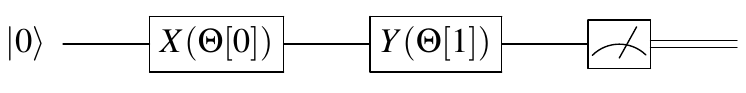}
	\caption{Single-qubit variational circuit $W(\Theta)$.\label{fig:quantumcircuit}}
\end{figure}

For this example, we use the variational quantum circuit pictured in Figure~\ref{fig:quantumcircuit}.
The input of the circuit is ground state $\ket{0}$.
Two rotation gates and a measurement gate are used.
The circuit consists of two quantum gates:
an $X$ gate and an $Y$ gate,
parameterized with rotations $\Theta[0]$ and $\Theta[1]$ about the $x$-axis and $y$-axis, on the Bloch sphere.
There is a measurement gate at the very end, converting the output qubit into a classical binary value.
This value is an action index.
The variational circuit is tuned by training such that it outputs the probably most rewarding  choice.

\begin{figure}[!htb]
    \centering

    \subfigure[Adversary is behaving arbitrarily.]{
        \includegraphics[width=0.4\columnwidth]{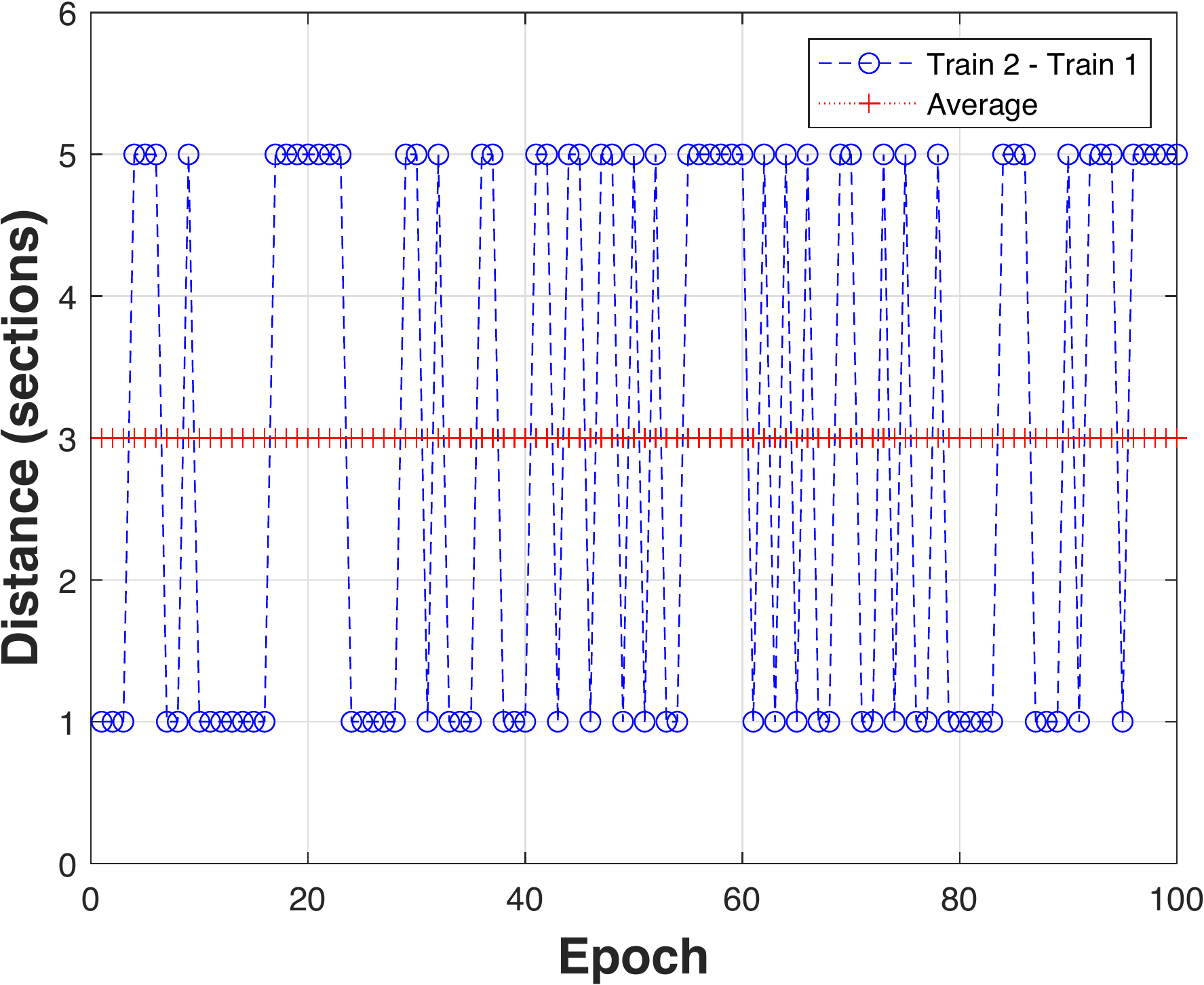}
    }
    \subfigure[Adversary mimics agent.]{
         \centering
         \includegraphics[width=0.4\columnwidth]{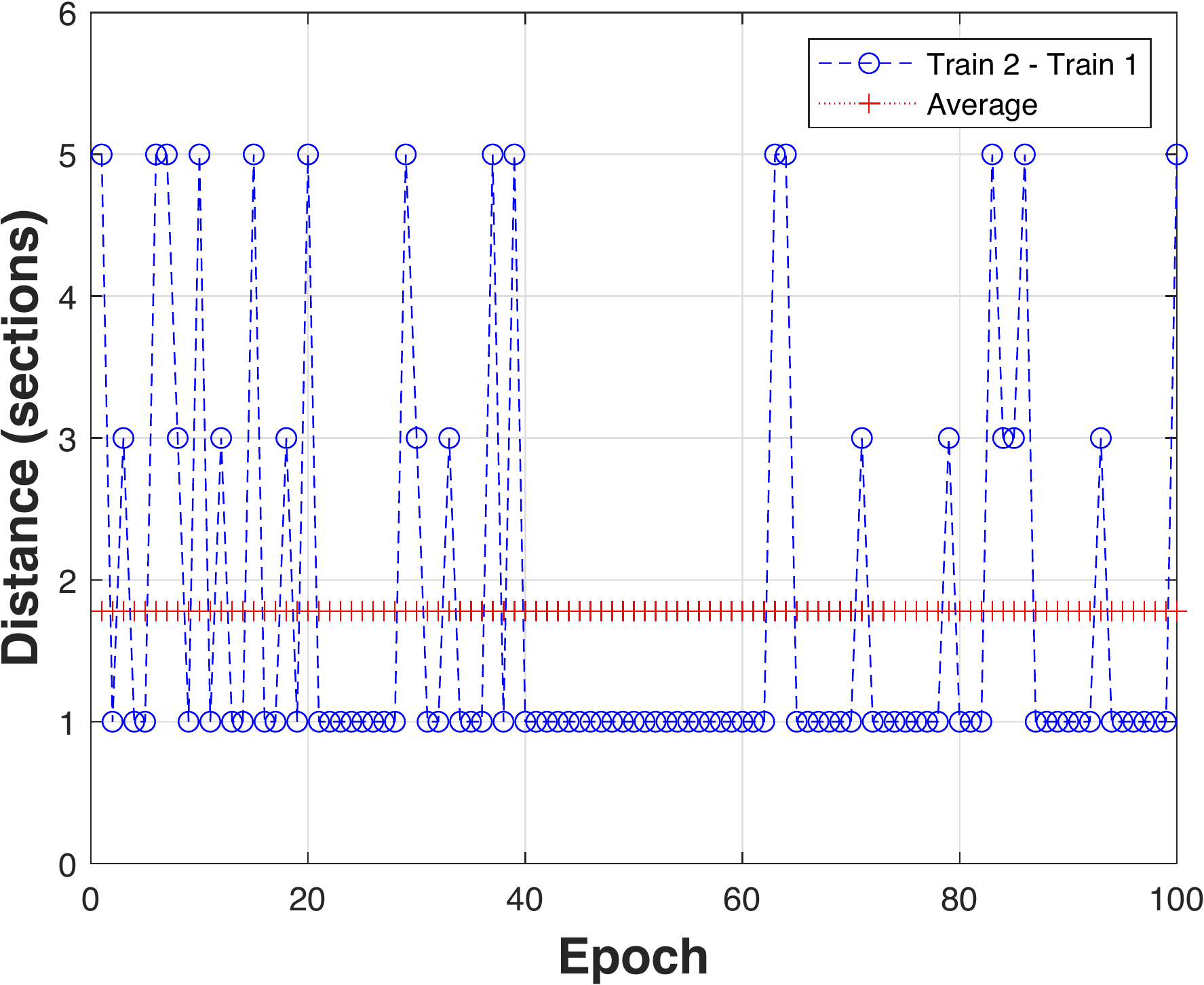}
    }
    \subfigure[Value of state zero.]{
       \includegraphics[width=0.4\columnwidth]{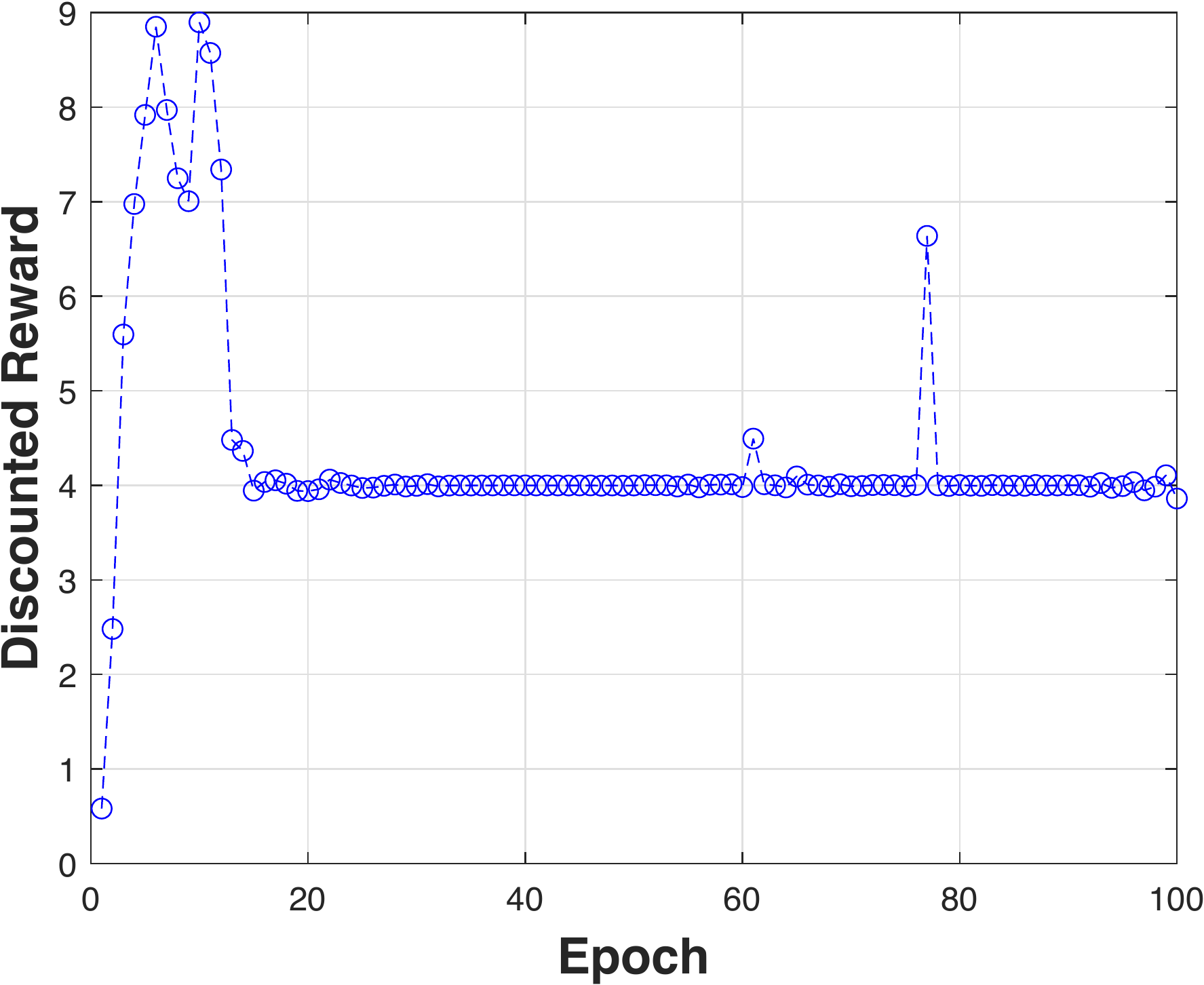}
    }
    \subfigure[Evolution of probabilities.]{
       \includegraphics[width=0.40\columnwidth]{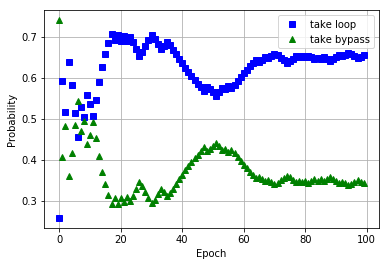}
    }
     \caption{(a)~The adversary randomly alternates between take loop and take bypass, with equal probabilities. (b)~The agent choices are leaking, e.g., due to the presence of an insider. With high probability, the adversary is mimicking the agent. (c)~Evolution of the value of state zero. (d)~Evolution of quantum variational circuit probabilities, with learning rate $\alpha$ is 0.01. \label{fig:results}}
\end{figure}

A detailed implementation of the example is available as
supplementary material in a companion github repository~\cite{github}. Figure~\ref{fig:results} provides graphical interpretations of the two-train example.
In all the plots, the $x$-axis represents epoch (time).
Part~(a) shows the Train~2-Train~1 separation distance (in sections) as a function of the epoch,
when the agent is doing the normal behavior, i.e., do action take loop, and the adversary is behaving arbitrarily, $p$ and $q$ are equal to $0.5$.
The average distance (three sections) indicates that more often the separation distance constraint is not violated.
Part~(b) also shows the Train~2-Train~1 distance as a function of the epoch,
but this time the adversary figured out the behavior of the agent.
The average distance (less than two sections) indicates that  the separation distance constraint is often violated.
Part~(c) plots the value of state zero, in Figure~\ref{fig:mdp}, versus epoch.
The adversary very likely learns the choices made by the agent, when at point~0. There is an insider leaking the information.
Train~2 is likely to mimic Train~1.
The probabilities of $p$ and $q$ are equal to $0.9$.
In such a case, for Train~1 the most rewarding choice is to take the loop.
Part~(d) shows the evolution of the probabilities of the actions, as the training of the quantum variational circuit pictured in Figure~\ref{fig:quantumcircuit} progresses.
They evolve consistently with the value of state zero (learning rate~$\alpha$ is~0.01).
The $y$-axis represents probabilities of selecting the actions take~loop (square marker) and
take bypass (triangle marker).
Under this condition, quantum \gls*{rl} infers that the maximum reward is obtained selecting the
\textit{take~loop} action. It has indeed higher probability than the
\textit{take~bypass} action.

\section{Discussion}
\label{sec:discussion}

Section~\ref{sec:example} detailed an illustrative example.
Of course, it can be enriched.
The successors of states~1 and~2 can be expanded.
More elaborate railways can be represented.
More sophisticated attack models can be studied.
For example,
let $v_i$ denote the velocity of Train~$i$, where
$i$ is equal to~1 or~2.
Hijacking control signals,
the adversary may slowly change the velocity of one of the trains
until the separation distance is not greater than a threshold $\tau$.
Mathematically, the velocity of the victimized train is represented as
$$
v_i (t+\Delta) = v_i(t) + \alpha e^{\beta (t+\Delta)}.
$$
The launched time of the attack is $t$.
$v_i(t)$ is the train velocity at time $t$,
while $v_i(t+\Delta)$ is the speed after a delay $\Delta$.
Symbols $\alpha$ and $\beta$ are constants.
During normal operation, the two trains are moving at equal constant velocities.
During an attack on the velocity of a train,
the separation distance slowly shrinks down to a value equal to or lower than a threshold.
The safe-distance constraint is violated.
While an attack is being perpetrated,
the state of the system must be recoverable~\cite{weerakkody2019resilient}, i.e.,
the velocities and compromised actuators or sensors can be determined using redundant sensing resources.

The approach can easily be generalized to other applications.
For instance, let us consider switching control strategies used to mitigate \gls*{dos} attacks~\cite{Zhu2020} or
input and output signal manipulation attacks~\cite{segovia2020}.
States are controller configurations, actions are configuration-to-configuration transitions and rewards are degrees of attack mitigation.
The variational circuit is trained such that the agent is steered in an attack mitigation condition.
This steering strategy is acquired through \gls*{rl}.

\begin{table}[htb]
\begin{center}
\caption{Conceptual comparison of classical versus quantum RL.}\label{tab:comparison}
\begin{tabular}{|c|c|c|}
\cline{1-3}
\multicolumn{3}{|c|}{Reinforcement Learning}
\\ \hline
Concept & Q-learning & Quantum \\ \hline
Data structure & Q-value table & Variational circuit \\
\hline
Resources & $n$ times $m$ numbers & $k\log n$ gates\\
\hline
\end{tabular}
\end{center}
\end{table}

In Section~\ref{sec:qrl},
quantum \gls*{rl} is explained referring to Q-learning.
Table~\ref{tab:comparison} compares Q-learning and quantum \gls*{rl}.
The first column list the  \gls*{rl} concepts.
The second column define their implementation in Q-learning~\cite{lecun2015deep}.
The third column lists their analogous in \gls*{qml}.
The core concept is a data structure used to represent the expected future rewards for action at each state.
Q-learning uses a table while \gls*{qml} employs a variational circuit.
The following line quantifies the amount of resources needed in every case.
For Q-learning, $n$ times $m$ expected reward numbers need to be stored, where $n$ is the number of states and $m$ the number of actions of the MDP.
For \gls*{qml},
$k \log n$ quantum gates are required, where $k$ is the number of gates used for each variational circuit line.
Note that  deep learning~\cite{lecun2015deep,mnih2015human,mnih2016asynchronous} and \gls*{qml} can be used to approximate the Q-value function,
with respectively, a neural network or a variational quantum circuit.
The second line compares tuneable parameters, which are neural network weight for the classical model and variational circuit rotations for the quantum model.
For both models, gradient descent optimization method is used to tune iteratively the model, the neural network or variational circuit.
Chen et al.~\cite{Chen2020} did a comparison of Deep learning and quantum \gls*{rl}.
According to their analysis, similar results can be obtained with similar order quantities of resources.
While there is no {\em neural network computer} in the works, apart for hardware~accelerators, there are considerable
efforts being deployed to develop the quantum computer~\cite{DigitaleWelt21}.
The eventually available quantum computer will provide an incomparable advantage to the ones who will have access to the technology, in particular the defender or adversary.

There are a few options for quantum encoding of states, including computational basis encoding, single-qubit unitary encoding and probability encoding.
They all have a time complexity cost proportional to the number of states.
Computational basis encoding is the simplest to grasp.
States are indexed $i=0,...,m-1$.
In the quantum format, the state is represented as $\ket{i}$.

Amplitude encoding works particularly well  for supervised machine learning~\cite{schuld2018supervised,barbeau2019recognizing}.
For example, let $\vec{\psi}=(\psi_0,\ldots,\psi_7)$ be such a unit vector.
Amplitude encoding means that the data is encoded in the probability amplitudes of
quantum states.
Vector $\vec{\psi}$ is mapped to the following three-qubit register
$$
\ket{\psi} = \sum_{i=0}^{7} \psi_i \ket{i}.
$$
The term $\ket{i}$ is one of the eight computational basis members for a three-qubit register.
Every feature-vector component $\psi_i$ becomes the probability amplitude of computational basis member $\ket{i}$.
The value $\psi^2$ corresponds to the probability of measuring the quantum register in state $\ket{i}$.
The summation operation is interpreted as the superposition of the quantum states  $\ket{i}$, $i=0,\ldots,7$.
Superposition means that the quantum state $\ket{\psi}$ assumes all the values of $i$  at the same time.
In this representation exercise, there is a cost associated with coding the feature vectors in the quantum format, linear in their number. The time complexity of an equivalent classical computing classifier is linear as well. However, in the quantum format the time taken to do classification is data-size independent.
The coding overhead, although, makes quantum \gls*{ml} performance comparable classical \gls*{ml} performance.
Ideally,  data should be  directly represented in the quantum format,
bypassing the classical to quantum data translation step and enabling gains in performance. Further research in quantum sensing is needed to enable this~\cite{Degen2017}.

There are also other \gls{rl} training alternatives.
Dong et al. have developed a quantum \gls*{rl} approach~\cite{Dong2008}.
In the quantum format, a state $i \in S$ of the \gls*{mdp} is mapped to quantum state $\ket{i}$.
Similarly, an action $j \in A$ is mapped to quantum state $\ket{j}$.
In state $i$, the action space is represented by the quantum state
$$
\label{eq:actionsuperposition}
\ket{A_i} = \sum_{j=0}^{m-1} \psi_i \ket{a_j}
$$
where the probability amplitudes $\psi_i$'s,
initially all equal, are modulated, using Grover iteration by the \gls*{rl} procedure.
In state $i$, selecting an action amounts to observing the quantum state $\ket{A_i}$.
According to the non-cloning theorem, it
can be done just once,  which is somewhat limited.

By far, not all \gls*{qml} issues have been resolved.
More research on encoding and training is required.
Variational circuit optimization experts~\cite{bergholm2020pennylane} highlight the need for more research to determine what works best,
among the available variational circuit designs,
versus the type of problem considered.

\section{Conclusion}
\label{sec:conclusion}

\noindent We have presented our vision of a next generation
cyber-physical defense in the quantum era. In the same way that nobody thinks about system protection making abstraction of the quantum threat, we claim that in the future nobody will think about  cyber-physical defense without using quantum resources.
When available, adversaries will use quantum resources to support their strategies.
Defenders must be equipped as well with the same resources to face 
quantum adversaries and achieve security beyond breach.
\gls*{ml} and quantum computing communities
will play very important roles in the design of such resources.
This way, the quantum advantage will be granted to defenders
rather than solely adversaries.
The essence of the war between defenders and adversaries is knowledge.
\gls*{rl} can be
used by an adversary for the purpose of system identification,
an enabler for covert attacks.
The paper has clearly demonstrated the plausibility of using quantum technique to search defense strategies and counter adversaries.
Furthermore, the design of new defense techniques can leverage quantum \gls*{ml}
to speedup decision making and support adaptive control.
These benefits of \gls*{qml} will although materialize when the quantum computer will be available.
These ideas have been explored in this article, highlighting capabilities and
limitations which resolution requires further research.\\
~~\\
\noindent \textbf{Acknowledgments ---} We acknowledge the financial support from the Natural Sciences and Engineering Research Council of Canada (NSERC) and the European Commission (H2020 SPARTA project, under grant agreement 830892).

\end{document}